\documentclass{article}
\usepackage{footnote}

\usepackage{arxiv}

\usepackage[utf8]{inputenc} 
\usepackage[T1]{fontenc}    
\usepackage{hyperref}       
\usepackage{url}            
\usepackage{booktabs}       
\usepackage{amsfonts}       
\usepackage{nicefrac}       
\usepackage{microtype}      
\usepackage{lipsum}		
\usepackage{graphicx}
\usepackage{cite}
\usepackage{doi}
\usepackage{amsmath}
\usepackage{babel}
\usepackage[ruled,lined]{algorithm2e}
\usepackage{xcolor}
\SetKwInput{KwOutput}{Output}  
\usepackage{multirow}
\usepackage{subcaption}
\usepackage{setspace} 
\title{Fast image reverse filters through fixed point and gradient descent acceleration}

\author{ 
    \href{https://orcid.org/0000-0000-0000-0000}{\includegraphics[scale=0.02]{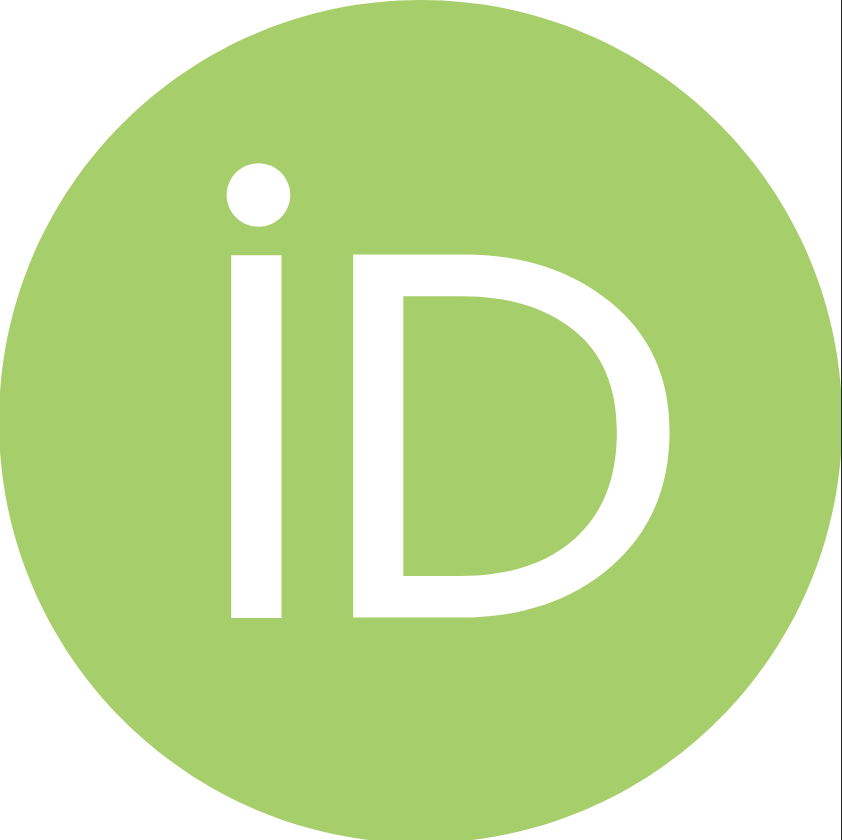}\hspace{1mm}Fernando J.~Galetto}\thanks{Corresponding author.} \\
  Department of  Engineering \\
  La Trobe University\\
  Bundoora, VIC 3086, Australia\\
	\texttt{f.galetto@latrobe.edu.au} \\
	\And
    \href{https://orcid.org/0000-0000-0000-0000}{\includegraphics[scale=0.02]{orcid.png}\hspace{1mm}Guang ~Deng}\\
    Department of  Engineering \\
    La Trobe University\\
    Bundoora, VIC 3086, Australia\\
    \texttt{d.deng@latrobe.edu.au} \\
}

\renewcommand{\shorttitle}{\textit{arXiv} Template}

\hypersetup{
pdftitle={A template for the arxiv style},
pdfsubject={q-bio.NC, q-bio.QM},
pdfauthor={David S.~Hippocampus, Elias D.~Striatum},
pdfkeywords={First keyword, Second keyword, More},
}

\begin{document}
\maketitle

\begin{abstract}

In this paper, we study the problem of reverse image filtering. An image filter denoted $g(.)$, which is available as a black box, produces an observation $\boldsymbol{b} = g(\boldsymbol{x})$  when provided with an input $\boldsymbol{x}$. The problem is to estimate the original input signal $\boldsymbol{x}$ from the black box filter $g(\boldsymbol{x})$ and the observation $\boldsymbol{b}$. We study and re-develop state-of-the-art methods from two points of view, fixed point iteration and gradient descent. We also explore the application of acceleration techniques for the two types of iterations. Through extensive experiments and comparison, we show that acceleration methods for both fixed point iteration and gradient descent help to speed up the convergence of state-of-the-art methods. 

\end{abstract}


\section{Introduction}

This paper presents an experimental study of accelerating  a class of semi-blind gradient free methods for solving an inverse problem in image processing. In terms of digital signal processing, solving such an inverse problem is commonly referred as  developing a reverse filter. The problem can be stated as follows. A filter denoted $g(\boldsymbol{x})$ is given as a black box of which the user can provide an input $\boldsymbol{x}$ and observe an output $\boldsymbol{b} = g(\boldsymbol{x})$ without knowing exactly how the filter produces certain effects such as edge-aware smoothing. The problem is to estimate the original input signal $\boldsymbol{x}$ from the black box filter $g(\boldsymbol{x})$ and the observation $\boldsymbol{b}$. Since the filter is unknown and is potentially highly nonlinear, traditional methods such as image restoration techniques \cite{Gonzalez_DIP}, which model the filter as a linear shift invariant system and try to estimate both the signal and the system, cannot be applied.  

This problem has been studied by researchers in recent years \cite{Tmethod},\cite{Fmethod},\cite{Rmethod},\cite{Pmethod},\cite{Deng_EL2019}, \cite{TDAmethod}. Methods that have been developed to solve this problem are based on formulation of either a fixed point iteration or a minimization of a cost function using gradient descent. In both formulations, the solutions are semi-blind in the sense that although they do not assume any knowledge of the filter, but they do rely on the input-output relationship of the filter. For the gradient descent based methods, a common feature is the of approximation of the the gradient of the cost function which involves the unknown filter function $g(.)$. These methods are called gradient-free, because they do not use required the exact gradient information, which are not available. 

One of the earliest method of semi-blind gradient free in literature is called zeroth order reverse filter \cite{Tmethod}, the method introduced a simple and efficient scheme justified by fixed-point iteration called Zero-Order Reverse filtering (T-method). A strict condition to guarantee convergence limits this method’s application to remove slight filtering effects only. Milanfar \cite{Rmethod} published a similar method here called R-method in which was obtained by minimizing a cost function composed of the cross-correlation and a term to avoid trivial solutions. The results do not show improvement against severe filtering effects compared with T-method. Dong \emph{et al.} \cite{Fmethod} presented an algorithm in the frequency domain called F-method. They based the method on the Newton-Raphson technique and focused only on convolution-based linear filters. The authors approximated the gradient using the forward difference to avoid getting the Jacobian of $g(.)$ . Although it reported better performance than past methods \cite{Tmethod, Rmethod}, it has two major limitations: it only deals with linear filters and it is unstable when the filter’s frequency response has zeros in the frequency range of $(0,\pi)$. Finally, in a recent work, Belyaev and Fayolle \cite{Pmethod} presented two methods based on Polak \cite{polyak1969minimization} and Steffensen \cite{steffensen1933remarks} gradient descent studies. Both deliver successful results and converge larger number of linear and non-linear filters than previous methods, but the 2-norm of a matrix needs to be calculated at each iteration, adding considerable computational cost to the process.

In our recent work\cite{TDAmethod}, we proposed a more efficient method called TDA-method which is a gradient free method justified trough total derivative approximation and gradient descent. Also we have explored the acceleration of the T- TDA-, and P-methods using some well known accelerated gradient descent (AGD) methods  \cite{optimizationAlgo} such as momentum (MGD) Nesterov (NAG), RMSProp, Adadelta and  Adam. 

A major motivation of this work is to further explore techniques and their performance in accelerating these gradient-free reverse filters. Since the T-method can be derived from solving a fixed-point problem point of view, a natural direction of study is to investigate the application of techniques for accelerating fixed-point iteration in these class of reverse filtering methods.

We note that numerical techniques are well established in accelerating fixed-point iterations through sequence transformation or extrapolation \cite{WENIGER1989189}\cite{SIAM-100-Digit}\cite{Numerical-Recipes}. We refer to reference \cite{iterativeFP_Acc} for a review of algorithms dealing with fixed point problems involving both scalar and vector variables. Reference \cite{Brezinski2000} provides an historical account for the main ideas and development of acceleration techniques based on sequence transformation.  A further motivation of this work comes from increasing research in applying such techniques in accelerating machine learning algorithms, which include techniques named after Anderson \cite{LupoPasini2021StableAA}\cite{Zhang2020GloballyCT}\cite{mai2020anderson}, Aitken \cite{Aitken2010}\cite{Chen2021RobustSG}, and Chebyshev \cite{Cheby-IEEE-SP-Lett}\cite{Anderson_cheby}.

The aim of this work is to study application of acceleration techniques for the two types of iterations: fixed point and gradient descent. We want to determine which acceleration technique works best (in terms of performance and computational complexity) which inverse filtering method and under what conditions such as the type of filtering effect to be inverted. In order to use fixed-point acceleration techniques, we also need to re-develop the TDA and P-method as fixed point iterations. They are originally derived from gradient descent point of view \cite{TDAmethod},\cite{Pmethod}.

The main contributions of this work can be summarized as the following. The application of acceleration techniques for fixed point iterations in this particular problem is new and have not been reported in literature. We have adopted a version of the Anderson acceleration \cite{anderson1965iterative} (Section \ref{sec:anderson-acceleration}) and also proposed a new way to define the Chebyshev sequence \cite{Cheby-IEEE-SP-Lett} (Section \ref{sec:Chebyshev}). We have also tested two vector variable acceleration methods which are related to Aitken's method \cite{aitken1925bernoulli}: Irons \cite{irons1969version} and Epsilon \cite{wynn1962acceleration} (Section \ref{sec:aitken_epsilon}) which are less well known in the image processing community. Other  contributions are as follows.  (1) A re-development of T-, TDA- and P-method from fixed point iteration point of view, which leads to another version of the P-method which is called p-method for easy reference. (2) An extensive evaluation of the 4 fixed point acceleration techniques as well as 4 gradient descent acceleration techniques. 

The organization of this paper is the following. In Section  \ref{sec:bk_info}, we present a brief previous works in gradient-free reverse filters and a brief introduction to the fixed point acceleration techniques used in this work. For completeness, we have also included acceleration techniques in this section.   In section \ref{sec:fixed-point-reverse-filter} we present  the formulation of the T-, TDA- and P-methods from fixed point iteration point of view. This is a new contribution of this paper. In Section \ref{sec:Results}, we present extensive experiments evaluations of 4 fixed point acceleration techniques and 4 gradient descent acceleration techniques for the reversing of effects from 14 filters which cover a wide spectrum of commonly used image processing techniques. These filters are summarized in Table \ref{tab:filters_parameters}.

\section{Reversing filters as fixed point iterations}\label{sec:bk_info} 

\subsection{A review of existing gradient-free reverse filters}
We follow the terminology used in \cite{Pmethod} which called a particular reverse filter a method and briefly comment on each of them in this section. We present a summary of related previous work in Table \ref{tab:compareAlgo} where we define the following variables which are used in the table and throughout this paper. 

\begin{equation}
    e(\boldsymbol{x}) = \boldsymbol{b} - g(\boldsymbol{x})
\end{equation}
\begin{equation}
    p(\boldsymbol{x}) = g(\boldsymbol{x}+e(\boldsymbol{x}))-g(\boldsymbol{x}-e(\boldsymbol{x}))
\end{equation}
\begin{equation}
    t(\boldsymbol{x})=g(\boldsymbol{x}+e(\boldsymbol{x}))-g(\boldsymbol{x})
\end{equation}

\begin{table}[h!]
\renewcommand{\arraystretch}{1.25}
    \centering
    \caption{A summary of reverse filter algorithms including the one calle "p-method" which is  derived from fixed point iteration in this paper. We present the reverse filtering from both fixed point iteration and gradient descent point of views. The function denoted $f(\boldsymbol{x})$  is key element in fixed point iteration, while the cost function denoted $c(\boldsymbol{x})$ is central to gradient descent. The complexity is measured in one iteration. The constant $C$ represents the complexity of the filter $g(.)$. The symbol $||\boldsymbol{x}||$ represents the matrix 2-norm which is the largest eigen value of the matrix $\boldsymbol{x}$. The symbol $\mathcal {F} $ represents the Fourier transform operator. }
    \label{tab:compareAlgo}

   \begin{tabular}{llllll}

\toprule

 \textbf{Method} & \textbf{Fixed point} (Section \ref{sec:fixed-point-reverse-filter}) & \textbf{Approx. gradient} \cite{TDAmethod}  &\textbf{Complexity} \tabularnewline
 & $f(\boldsymbol{x})$  & $-\nabla c(\boldsymbol{x})$  &   &\tabularnewline
\midrule

T \cite{Tmethod}  & $f_T(\boldsymbol{x})=\boldsymbol{x}+ \lambda e(\boldsymbol{x})$ & $e(\boldsymbol{x})$  & $\max{(O(n),C)}$ \tabularnewline

R \cite{Rmethod} & $f_R(\boldsymbol{x})=\alpha \boldsymbol{x}+\lambda  e(\boldsymbol{x})$ & $e(\boldsymbol{x})$ & $\max{(O(n),C)}$\tabularnewline

P  \cite{Pmethod} & $f_P(\boldsymbol{x})=\boldsymbol{x}+\frac{||e(\boldsymbol{x})||}{2||p(\boldsymbol{x})||} p(\boldsymbol{x})$ & $\frac{1}{2}p(\boldsymbol{x})$ & $\max{(O(n^2),C)}$ \tabularnewline

TDA \cite{TDAmethod}& $f_{TDA}(\boldsymbol{x})=\boldsymbol{x}+\lambda t(\boldsymbol{x})$  & $t(\boldsymbol{x})$  & $\max{(O(n),C)}$ \tabularnewline

F \cite{Fmethod} & $f_F(\boldsymbol{x}) = \mathcal{F}^{-1} \left(  \frac{\mathcal{F}(\boldsymbol{b}) }{\mathcal{F}(g(\boldsymbol{x}_k))\mathcal{F}(\boldsymbol{x}_k)} \right ) $  & N/A & $\max{(O(n\log n),C)}$ \tabularnewline

p (this work) &  $f_p(\boldsymbol{x})=\boldsymbol{x}+ \frac{1}{2}p(\boldsymbol{x})$ &  $\frac{1}{2}p(\boldsymbol{x})$ & $\max{(O(n),C)}$  \tabularnewline

\bottomrule 
\end{tabular}
\label{tab:method_summary}
\end{table}

In this table, we present these methods from both the fixed point iteration (detailed in Section \ref{sec:fixed-point-reverse-filter}) and gradient descent (detailed in our previous work \cite{TDAmethod}) point of views. In Section \ref{sec:implement_relation}, we briefly discuss the implementations and relationships between the two point of views based on this table. To make the discussion of this paper self contained, we present a brief review of each method by pointing out its main idea and advantages/limitations. The review is similar to that presented in our previous work \cite{TDAmethod} and is included in Appendix A for easy reference.  

\subsection{Reverse filter as fixed-point iterations}\label{sec:fixed-point-reverse-filter}

A fixed-point problem denoted by: $ \boldsymbol{x} = f(\boldsymbol{x}), f: \mathbb{R}^N \rightarrow \mathbb{R}^N $,  can be solved by the Picard iteration 
\begin{equation}\label{eq:picard_iteration}
\boldsymbol{x}_{k+1}=f(\boldsymbol{x}_{k})
\end{equation}
Starting with an initial guess $x_0$, the iteration is performed until a converge criterion is met. Fixed point iterations may present two type of convergence problems: a) the iteration may never converge, b) the iterates presents linear convergence which can be very slow, specially if the evaluation of $f(x)$ is computationally expensive\cite{walker2011anderson}. We remark that there are other type of iterations such as Mann iteration \cite{MannIteration}, segmenting Mann iteration \cite{FPiteration4RealFunctions} which is the same form as the  over relaxation method \cite{Cheby-IEEE-SP-Lett} defined as
\begin{equation} \label{eq:Mann}
   \boldsymbol{x}_{k+1}= \boldsymbol{x}_k+ \omega_k \left ( f(\boldsymbol{x}_k)-\boldsymbol{x}_k \right )   
\end{equation}
where $\omega_k$ is a real coefficient. In this work, we focus on the Picard iteration the over relaxation method as candidates for accelerations.
We first use the T-method as an example to review the basic idea of fixed point iteration. We then derive TDA- and P-methods from a fixed point iteration point of view. 
\subsubsection*{T-method}\label{sec:fp-T-method}
  We recall that the T-method can be derived by rewriting the filter model $\boldsymbol{b} = g(\boldsymbol{x})$ as a fixed-point problem in which 
\begin{equation} \label{eq:T-iteration}
\boldsymbol{x} = f_T(\boldsymbol{x})=\boldsymbol{x} +e(\boldsymbol{x})
\end{equation}
where we use $f_T$ to indicate the function is for the T-method. 

\subsubsection*{TDA-method}

Following the T-method, we can derive the TDA-method as a fixed-point problem by applying the filter $g$ to both sides of equation (\ref{eq:T-iteration}), adding  $\boldsymbol{x}$ to both sides, and moving $g(\boldsymbol{x})$ to the right hand side

\begin{align}
\boldsymbol{x} &= f_{TDA}(\boldsymbol{x}) \\
&= \boldsymbol{x}+g(\boldsymbol{x}+e(\boldsymbol{x}))-g(\boldsymbol{x})
\end{align}

\subsubsection*{The p-method as fixed point iteration}

We use a small letter "p" to distinguish between the one developed in this section and the one developed in reference \cite{Pmethod}.  When $\boldsymbol{x}$ is the fixed point we have $e(\boldsymbol{x}) = 0$. Thus, we rewrite the filter model  as 

\begin{equation}
\boldsymbol{x} - e(\boldsymbol{x})=\boldsymbol{x} + e(\boldsymbol{x})
\end{equation} 
This equation can be further rewritten as 
\begin{equation}
\boldsymbol{x} + \frac{1}{2}g(\boldsymbol{x} - e(\boldsymbol{x})) = \boldsymbol{x} + \frac{1}{2}g(\boldsymbol{x} + e(\boldsymbol{x}))
\end{equation}
where the scaling factor 1/2 is introduced such that a connection between the p-method and the P-method can be developed. Re-arranging, we have derived the following fixed point problem
\begin{align}
\boldsymbol{x} &= f_p(\boldsymbol{x}) \\
&= \boldsymbol{x} +  \frac{1}{2}(g(\boldsymbol{x} + e(\boldsymbol{x}))  - g(\boldsymbol{x} - e(\boldsymbol{x})))
\end{align}

 Comparing the p-method with the P-method, we can see that the former is without the computational expensive factor $\frac{||\boldsymbol{q}_{k}||}{2||\boldsymbol{p}_{k}||}$. On the other hand, since the P-method can be regarded as an approximation of the gradient descent, the p-method can be similarly interpreted because they share the same form.

\subsection{Implementations and relationships}\label{sec:implement_relation}
Referring to Table \ref{tab:method_summary}, we can implement the reverse filter methods either as a Picard iteration using equation (\ref{eq:picard_iteration}) or the segmenting Mann iteration using equation (\ref{eq:Mann}). An interesting connection between the segmenting Mann iteration and the gradient descent can be established by relating the two implementations as follows. Let us use the TDA-method as an example, the implementations of gradient descent and segmenting Mann iteration can be written as

\begin{align}\label{eq:GD-}
    \boldsymbol{x}_{k+1} &= \boldsymbol{x}_k +\lambda_k (-\nabla c(\boldsymbol{x}_k)) \\
    &= \boldsymbol{x}_k +\lambda_k t(\boldsymbol{x}_k)
\end{align}
and
\begin{align}\label{eq:SOR}
\boldsymbol{x}_{k+1} &= \boldsymbol{x}_{k}+\omega_{k}(f_{TDA}(\boldsymbol{x}_{k}) - \boldsymbol{x}_{k}) \\
&=  \boldsymbol{x}_k +\omega_k t(\boldsymbol{x}_k)
\end{align}
This is not a surprising result. Indeed, from Table \ref{tab:method_summary}, we can verify that for the reverse filters (except the F-method), the follow relationship holds  
\begin{equation}\label{eq:relation-fp-GD}
    -\nabla c(\boldsymbol{x}) =  f(\boldsymbol{x}) -\boldsymbol{x}   
\end{equation}

In addition, the relationship between the derivation from both gradient descent and fixed-point can be established from the following perspective. From an optimization point view, estimating the unknown image from the observation $\boldsymbol{b}$ and a black box filter $g(\boldsymbol{x})$ can be formulated as solving a minimization problem with the cost function $c(\boldsymbol{x})$. Assume there is a least a local minimum $\bar{\boldsymbol{x}}$ which satisfies the condition: $-\nabla c(\bar{\boldsymbol{x}})=0$. The optimization problem becomes one that solves a system of nonlinear equations:  $-\nabla c(\boldsymbol{x})=0$. One way to solve this problem is to formulate it as a fixed point iteration writing: 
\begin{equation}
    \boldsymbol{x}=\boldsymbol{x}-\nabla c(\boldsymbol{x}) 
\end{equation}
In terms of definition given by equation (\ref{eq:picard_iteration}), we have $f(\boldsymbol{x})=\boldsymbol{x}-\nabla c(\boldsymbol{x})$. We then perform the segmenting Mann iteration by substituting $f(\boldsymbol{x})$ into equation (\ref{eq:Mann}), we have the same equation as that of the gradient descent. Therefore, a reverse filter based on the segmenting Mann iteration is equivalent to the same one based of the gradient descent. We remark that it is well known that gradient descent (with a fixed parameter) can be regarded as an fixed point iteration when we define $f(\boldsymbol{x}_k)=\boldsymbol{x}_k-\lambda_k \nabla c(\boldsymbol{x}_k)$  where $\lambda_k$ is a constant. In this section, we specifically point out equivalence of the segmenting Mann iteration and gradient descent, which allows us to explore the application of accelerations techniques from both gradient descent and fixed-point literature.        

\section{Acceleration techniques}
\subsection{Acceleration of fixed point iterations}
 In the following, we review four acceleration methods: Chebyshev periodical successive over-relaxation \cite{Cheby-IEEE-SP-Lett}, Anderson acceleration \cite{anderson1965iterative}, Irons acceleration \cite{irons1969version} and Wynn's $\epsilon$-algorithm \cite{wynn1962acceleration}. 

\subsubsection{Chebyshev periodical successive over-relaxation}\label{sec:Chebyshev}
In essence, the Chebyshev periodical successive over-relaxation (PSOR)  \cite{Cheby-IEEE-SP-Lett} can be regarded as another method for the calculation of the coefficient in equation (\ref{eq:Mann}).  The Chebyshev sequence is periodic with a period of $T$ and is defined as:
\begin{equation}
\omega_k = \left [ \frac{\lambda_2+\lambda_1}{2} + \frac{\lambda_2-\lambda_1}{2} \cos{\left ( \frac{2k+1}{2T} \pi \right )}   \right ]^{-1}
\end{equation}
where  $\lambda_2$ and $\lambda_1$ are the lower and upper bound of the sequence and are user defined parameters. It has been demonstrated that such method can help acceleration the convergence of fixed point iteration  \cite{Cheby-IEEE-SP-Lett}. A recent paper also showed that using the Chebyshev sequence as the coefficients for Anderson acceleration leads to good results in machine learning applications \cite{Anderson_cheby}. In reference \cite{cheby2020}, it was shown that using the settings of $\lambda_1=0.18$, $\lambda_2=0.98$ and $T=8$ leads to good acceleration of the T-method in reversing the effect of mild low pass filter operation on a low resolution small hand-written images. 

In our experiments we observed that a direct application of the above Chebyshev sequence to the reverse filters did not produce satisfactory results. Thus, we experimented with other parameter settings. of. We found that when we set $\lambda_1=0$ and $\lambda_2=1$ and clip the sequence is an upper limit $\alpha$ as defined in equation (\ref{eq:new_cheby}), all inverse filters can produce consistent good results over a broad spectrum of filters.
\begin{equation}\label{eq:new_cheby}
\omega_k = \min\left(\alpha, \frac{2}{ 1 +  \cos\left ( \frac{2k+1}{2T} \pi \right )}\right )
\end{equation}

It is important to study the value of the Chebyshev sequence, because it directly determines whether or not the iteration would converge. For a detailed study of this topic, we refer to reference \cite{FPiteration4RealFunctions}. For a filter function $f$ which is L-Lipschitz, the convergence condition is $\omega_k\le\frac{2-\delta}{1+L}<2$ for some $\delta>0$ (Proposition 6, \cite{FPiteration4RealFunctions}). In light of this convergence condition, we have the following remarks about the Chebyshev sequence in terms of the period $T$.
\begin{itemize}
\item  Within one period it is nonlinearly increasing with $k$. The minimum and maximum can be calculated as 
\begin{equation}
 \min\{\omega_{k}\}=\omega_{0} =\frac{2}{1+\cos\left(\frac{\pi}{2T}\right)}
\end{equation}
 and 
\begin{equation}
\max\{\omega_{k}\}=\omega_{T-1}=\frac{2}{1-\cos\left(\frac{\pi}{2T}\right)}
\end{equation}
Since both $\omega_{0}$ and $\omega_{T-1}$ are a function of $T$, we can determine how the setting of $T$ will change lower and upper bound of the sequence. Because $\cos\left(\frac{\pi}{2T}\right)$ is an increasing function of $T$ and $\cos\left(\frac{\pi}{2T}\right)\rightarrow1$ when $T>>1$. Thus we have $1\le\omega_{0}<\omega_k$. Similarly, we can see that the maximum value of $\omega_{T-1}$ is unbounded when $T>>1$. In fact $\omega_{T-1}$ is nonlinearly increasing with $T$. A bigger $T$ value leads to
a bigger value of $\omega_{T-1}$.
\item Thus, setting a bigger $T$ value has three effects.
\begin{itemize}
\item Increasing the number of iterations in which $\omega_{k}>1$ before
it resets to $\omega_{0}$.
\item Increasing the value of $\omega_{k}$ when $k$ is closed to $T$.
\item Decreasing $\omega_{0}$ to near 1.
\end{itemize}
\item In one period, there can be a number of $\omega_k$'s whose values are greater than 2. Let this number be $N_T$ which is controlled by $T$. A bigger $T$ value will result in a bigger $N_T$. Within one period, the role of $N_T$ iterations with $\omega_k>2$ (at the end of the period) is to push the result out of a potential local minimum \cite{loshchilov2016sgdr}. In fact, the use of the Chebyshev sequence in accelerating the segmenting Mann iteration is closely related to the so-called cosine-annealing with warm start\footnote{Such strategies are widely used in adapting the learning rate of gradient descent. See for example implementation in Pytorch: https://pytorch.org/docs/stable/generated/torch.optim.lr\_scheduler.CosineAnnealingLR.html}. However, we observed in our experiment that when $N_T$ is too large, the result is pushed too far way from the optimum leading to sub-optimal results or non-stable iteration. One way combat this problem is to properly choose $T$ such that $N_T$ is small and to clip the value of $\omega_k$ to a predefined value $\alpha$.  In our experiments, we set $T=32$ and $\alpha=3$  for T-, TDA- and p-Method. We set $\alpha=1$ for the P-method.  
\end{itemize}

\subsubsection{Anderson Acceleration}\label{sec:anderson-acceleration}
Anderson acceleration (AA), also known as Anderson interpolation was first proposed in \cite{anderson1965iterative} and deeper discussed in \cite{walker2011anderson} and \cite{anderson2019comments}.  The method  redefines $x_{k+1}$ to use not just the last iterate but also $m_k$ samples of past information to generate a new estimation.
\begin{equation}
\boldsymbol{x}_{k+1} = f(\boldsymbol{x}_k) - \sum_{j=0}^{m_k-1} \theta_j^k (f(\boldsymbol{x}_{k-m_k+j+1})-f(\boldsymbol{x}_{k-m_k+j})) 
\end{equation}
where $\theta$ values can be found trough solving a minimization problem, which following the steps in \cite{toth2015convergence, walker2011anderson, chen2022non} can be optimally reformulated for implementation as an unconstrained linear least squared problem :
\begin{equation}
\min_{\theta_1 ... \theta_{m_k}}||F(\boldsymbol{x}_k) + \sum_{j=0}^{m_k-1} \theta_j^k (F(\boldsymbol{x}_{k-m_k+j+1})-F(\boldsymbol{x}_{k-m_k+j})) ||_2^2
\end{equation}
where $F(\boldsymbol{x}) = f(\boldsymbol{x}) - \boldsymbol{x}$. The algorithm's implementation does not require derivatives which makes it ideal to accelerate reverse filtering methods.  

\subsubsection{Acceleration of matrix variable based on generalization of scalar variable algorithms}\label{sec:aitken_epsilon}
Both Aitken's method and the $\epsilon$-algorithm were originally proposed for scalar variable case. A limitation is that the value of the denominator in both methods can be potentially very small or even zero which makes the iteration unstable. In our experiment with images, this happens quite frequently. The vector variable version is more stable, because there is less chance that the value of the denominator $||\Delta^2(\boldsymbol{x}_k)||^2$ is very small.  In the following, we first review the main ideas of generalization of acceleration techniques from scalar variable to vector variable. We then describe the proposed generalization to matrix variables based on such ideas. 

We consider the Irons' method and $\epsilon$ method which can be regarded as generalization of 
Aitken's $\Delta^2$ process  \cite{aitken1925bernoulli}. Aitken's method was designed to accelerate the convergence of scalar sequences generated by an iterative process. Steffensen \cite{steffensen1933remarks} later applied recursively to solve fixed point problems such $x=f(x)$ where $x \in \mathbb{R} $ and $ f: \mathbb{R} \rightarrow \mathbb{R} $. Specifically, Aitken's method can be written as
\begin{align}
    x_{k+1} &= x_k -\frac{(\Delta x_k)^2}{\Delta^2x_k} \\
            &= x_k - ((\Delta x_k)(\Delta^2x_k)^{-1})\times (\Delta x_k)  \label{eq:aitken}
\end{align}
where $\Delta x_k = f(x_{k})-x_k$, $\Delta f(x_k) = f(f(x_k)) - f(x_k)$, and $ \Delta^2x_k = \Delta f(x_{k}) - \Delta x_k$.

A vector variable generalization of Aitken's method can thus be written by replacing the scalar variable by vector variable in eqation (\ref{eq:aitken}) as follows
 \begin{equation} \label{eq:vectorAitken}
    \boldsymbol{x_{k+1}} = \boldsymbol{x_{k}} - ((\Delta \boldsymbol{x_{k}})^T(\Delta^2 \boldsymbol{x_{k}})^{-1})\Delta \boldsymbol{x_{k}} 
\end{equation}
The inverse of a vector called Samelson's inverse \cite{matrixTheory_Gentle}  is defined as
\begin{equation}
\boldsymbol{x}^{-1} = \frac{\boldsymbol{x}}{||\boldsymbol{x}||^2}
\label{eq:vector_inverse}
\end{equation}
where $||\boldsymbol{x}||$ is the Euclidean norm \footnote{We note that in Wynn's paper \cite{wynn1962acceleration}, the definition is $\boldsymbol{x}^{-1} = \frac{\boldsymbol{x}^T}{||\boldsymbol{x}||^2}$.}. Irons' method \cite{irons1969version} is a generalization of Aitken's method by replacing $(\Delta^2 \boldsymbol{x}_k)^{-1}$ in equation (\ref{eq:vectorAitken}) with its Samelson inverse. The vector version of Aitken's method is thus given by

\begin{equation} \label{eq:vectorAitken1}
    \boldsymbol{x_{k+1}} = \boldsymbol{x_{k}} - \frac{(\Delta \boldsymbol{x_{k}})^T \Delta^2 \boldsymbol{x_{k}}}{||\Delta^2 \boldsymbol{x_{k}}||^2} \Delta \boldsymbol{x_{k}} 
\end{equation}
The $\epsilon$-algorithm was first proposed by Wynn \cite{wynn1961epsilon}  \cite{wynn1962acceleration} as alternative to extrapolate a fixed point iteration. For a scalar fixed-point problem, the $\epsilon$-algorithm \cite{Morris-extrapolation-epsilon92} can be written as
\begin{align}
    x_{k+1} &= f(x_k) - \frac{\Delta x_k \Delta f(x_k) }{\Delta^2 x_k} \\
            &= f(x_k) - \frac{\Delta x_k |\Delta f(x_k)|^2 - |\Delta x_k|^2 \Delta f(x_k) }{|\Delta^2 x_k|^2}
\end{align}
where variables are defined in the same way as those in Aitken's method.
The generalization to the vector variable case \cite{Morris-extrapolation-epsilon92}  can be achieved by replacing the square absolute value with the vector norm and the result is the following
\begin{equation}
\boldsymbol{x}_{k+1} = f(\boldsymbol{x}_{k})+ \frac{|| \Delta \boldsymbol{x}_k ||^2 \Delta f(\boldsymbol{x}_k) - ||\Delta f(\boldsymbol{x}_k) ||^2  \Delta \boldsymbol{x}_k}{||\Delta^2\boldsymbol{x}_k||^2}
\end{equation}

We now present the proposed generalization to matrix variable case. For the $\epsilon$-algorithm, the generalization is straight forward. We just need to regard all variables in equation () as matrix replace the 
In each iteration, both Aitken's method and the $\epsilon$-algorithm require two calls of the original function $f(.)$ and the norm calculation.

\subsection{Acceleration of gradient descent}

Gradient descent is an iterative method for finding a point $\hat{\boldsymbol{x}}$ which minimizes the cost function $c(\boldsymbol{x})$. A general form of gradient descent is given by 
\begin{equation}
    \boldsymbol{x}_{k+1}=\boldsymbol{x}_k -\lambda \nabla c(\boldsymbol{x}_k) 
\end{equation}
Accelerated gradient descent (AGD) methods \cite{optimizationAlgo},  were developed to tackle problems related gradient descent, some of them are summarised in Table \ref{tab:AGD_table_summary}. They are widely used training deep neural networks and have the potential to reduce the number of iterations needed to achieve convergence and to increase the robustness against noisy gradients \cite{Goodfellow-et-al-2016}. Since the T-, TDA- and P-method can be regarded as gradient descent algorithms, as proved in our previous work \cite{TDAmethod} these AGD methods can be used to improve the performance of the three algorithms. The tests in this paper are conducted by replacing the gradient of each AGD method with the gradient  approximations shown in Table \ref{tab:compareAlgo} method summary.

\begin{table}
\centering
\caption{AGD methods.}
\label{tab:AGD_table_summary}
\renewcommand{\arraystretch}{1.5}
\begin{tabular}{ll}

\toprule
\textbf{Gradient descent}                        & $\boldsymbol{x}_{k+1}=\boldsymbol{x}_k -\lambda \nabla c(\boldsymbol{x}_k)$                                                                       \vspace{0.25cm}  \\ 
\textbf{MGD}                     & $\boldsymbol{x}_{k+1}=\boldsymbol{x}_k +\boldsymbol{v}_k $   \\
 & $ \boldsymbol{v}_k = \beta \boldsymbol{v}_{k-1} -  \lambda \nabla c(\boldsymbol{x}_k)$ \vspace{0.25cm}  \\ 
\textbf{NAG}                     & $\boldsymbol{x}_{k+1}=\boldsymbol{x}_k +\boldsymbol{v}_k $     \\  & $ \boldsymbol{v}_k = \beta \boldsymbol{v}_{k-1} -  \lambda \nabla c(\boldsymbol{x}_k + \beta \boldsymbol{v}_{k-1})$ \vspace{0.25cm}  \\ 

\textbf{RMSProp}                    & $\boldsymbol{x}_{k+1}=\boldsymbol{x}_k -\frac{\lambda}{\sqrt{\boldsymbol{v}_k+\epsilon}}\nabla c(\boldsymbol{x}_k)$                                \\ 
                                   & $\boldsymbol{v}_k = \beta \boldsymbol{v}_{k-1} +(1- \beta)(\nabla c(\boldsymbol{x}_k))^2$                                                           \vspace{0.25cm}  \\ 
\textbf{Adadelta} & $ \boldsymbol{x}_{k+1}=\boldsymbol{x}_k -\Delta \boldsymbol{x}_{k}$ \\
                                    & $\Delta \boldsymbol{x}_{k} =\frac{\sqrt{\boldsymbol{u}_{k-1}+\epsilon}}{\sqrt{\boldsymbol{v}_{k}+\epsilon}}\nabla c(\boldsymbol{x}_k)$                                        \\
                                    & $\boldsymbol{v}_{k} = \beta \boldsymbol{v}_{k-1}  + (1-\beta) (\nabla c(\boldsymbol{x}_k))^2$\\
                                   & $\boldsymbol{u}_{k} = \beta \boldsymbol{u}_{k-1}  + (1-\beta) (\Delta \boldsymbol{x}_{k})^2 $   \vspace{0.25cm}                                                     \\

                                \textbf{ADAM} & $\boldsymbol{x}_{k+1}=\boldsymbol{x}_k -\lambda \frac{\boldsymbol{\hat{m}_k}}{\sqrt{\boldsymbol{\hat{v}_k}}+\epsilon}$                             \\
                                   & $\boldsymbol{m_k} = \beta_1 \boldsymbol{m_{k-1}} + (1-\beta_1) \nabla c(\boldsymbol{x}_k)$  \\
                                   & $\boldsymbol{v_k} = \beta_2 \boldsymbol{v_{k-1}} + (1-\beta_2) (\nabla c(\boldsymbol{x}_k))^2$  \\
                                   & $\boldsymbol{\hat{m}_k} = \frac{\boldsymbol{m_k} }{1-\beta_1}$,
                                    $\boldsymbol{\hat{v}_k}= \frac{ \boldsymbol{v_k}}{1-\beta_2}$  
                        \vspace{0.25cm}  \\  \textbf{SGDR}\cite{loshchilov2016sgdr} &                 $\lambda_k = \lambda_{min} + \frac{1}{2} \left ( \lambda_{max} - \eta_{min} \right )\left ( 1+\cos\left(\frac{T_{cur}}{T}\pi\right) \right )$                             \\
                                   & $\boldsymbol{x}_{k+1}=\boldsymbol{x}_k - \lambda_k \nabla c_k(\boldsymbol{x}_k)$  \\
                         \bottomrule 
                                    
\end{tabular}
\end{table}

Since the methods in Table \ref{tab:AGD_table_summary} were studied in \cite{TDAmethod} and for the sake of simplicity we only use MGD, NAG and Adam in this paper. Also, in this paper we study the use of a periodic warm restart technique to improve gradient descent based reverse image filtering methods using the scheme proposed in \cite{loshchilov2016sgdr}. The idea is taken from gradient free optimization techniques which included restarts and schedule mechanisms for multi modal functions to avoid local optimums or minimums.

The essential difference between SGDR and the classic SGR is that the learning rate or step size 
$\lambda_k$ is not fixed. SGDR proposed a learning rate which is scheduled to change after each iteration, as shown in Eq.\ref{eq:SGDR}. The learning rate is initialized in the first iteration, then incremented after each iteration. The learning rate cycle repeats after $T$ number of iterations: 

\begin{equation}
    \lambda_k = \lambda_{min} + \frac{1}{2} \left ( \lambda_{max} - \eta_{min} \right )\left ( 1+\cos\left(\frac{T_{cur}}{T}\pi\right) \right )
    \label{eq:SGDR}
\end{equation}
where $\eta_{max}$ and $\eta_{min}$ are the limits for the learning rate, $T_{cur}$ indicates how many iterations passed since last restart. 

SGDR has the potential to have a faster convergency than SGD with static learning rate since $\eta_{max}$ can be set to be higher than the optimum learning rate for SGD.

\subsection{Computational complexity}

Reverse image filtering methods such as T, TDA, P and p methods have a computational complexity associated. When an acceleration is applied the complexity could be incremented, however, we expect the improvement in the performance to overcome the drawbacks of complexity. i.e accelerated methods should be able to achieve the same results with less number of iterations than non-accelerated methods. In this section we study acceleration methods and measure their computational complexity in one iteration, as well as the number of times $\#$ that the black box or filter $g(.)$ is required per iteration. A summary is shown in Table \ref{tab:complexity}, where $C$ represents the complexity of the filter. 

In the first row of Table \ref{tab:complexity}, we can see the original methods without acceleration. The P method is the most complex algorithm with $\max{(O(n),C)}$, it requires to calculate the 2-norm of a matrix which makes it very slow for high resolution images. ALso the P method requires 3 instances of the filter to be reverse per iteration. On the other hand, the T method is the most efficient algorithm since only requires 1 instance of the original filter and has a complexity of $\max{(O(n),C)}$. 

The rest of the table shows that GD based accelerations hold the complexity of the original method as well as the number of calls required per iteration. However, AFP methods such as EPSILON and IRONS increment the complexity since their require the calculation of the 2-norm and twice the number of filter calls than the original methods which is critical specially when accelerating the P method when the number of function calls jumps from 3 to 6.

\begin{table}[h!]
\centering
\caption{Computational complexity measured in one iteration for the 4 reverse image filtering methods and its accelerations where $C_1=\max{(O(n),C)}$ and $C_2=\max{(O(n^2),C)}$. The constant $C$ represents the complexity of the filter $g(.)$ and $\#$ indicates the number of filter calls $g(.)$ per iteration.}

\label{tab:complexity}
\begin{tabular}{|l|ll|ll|ll|ll|}
\toprule
& \multicolumn{2}{c|}{\textbf{T}} & \multicolumn{2}{c|}{\textbf{TDA}} & \multicolumn{2}{c|}{\textbf{P}} & \multicolumn{2}{c|}{\textbf{p}} \\
 \midrule
  Acceleration & CXTY        & $\#$      & CXTY         & $\#$          & CXTY        & $\#$        & CXTY        & $\#$        \\
 \midrule
\textbf{None} & $C_1$   & 1 & $C_1$    & 2 & $C_2$    & 3 & $C_1$    & 3 \\
\textbf{MGD}      & $C_1$    & 1 & $C_1$    & 2 & $C_2$    & 3 & $C_1$    & 3 \\
\textbf{NAG}      & $C_1$    & 1 & $C_1$    & 2 & $C_2$    & 3 & $C_1$    & 3 \\
\textbf{SGDR}     & $C_1$    & 1 & $C_1$    & 2 & $C_2$    & 3 & $C_1$    & 3 \\
\textbf{ADAM}     & $C_1$    & 1 & $C_1$    & 2 & $C_2$    & 3 & $C_1$    & 3 \\
\textbf{IRONS}    & $C_2$  & 2 & $C_2$  & 4 & $C_2$  & 6 & $C_2$  & 6 \\
\textbf{EPSILON}  & $C_2$  & 2 & $C_2$ & 4 & $C_2$  & 6 & $C_2$  & 6 \\
\textbf{AAcc}     & $C_1$   & 1 & $C_1$    & 2 & $C_2$    & 3 & $C_1$    & 3 \\
\textbf{CHB}      & $C_1$    & 1 & $C_1$   & 2 & $C_2$    & 3 & $C_1$    & 3 \\
\bottomrule
\end{tabular}
\end{table}

\section{Results and discussion}\label{sec:Results}

In this section we demonstrate the effectiveness of the proposed acceleration techniques to improve the convergence of the classic T, TDA, P and p method. First, in section \ref{sec:filter1} and \ref{sec:filter2} we reverse two typical image filters (1) Guided image filtering \cite{he2012guided}, and (2) Motion blur filter with different method and accelerations using a single image. Then in section \ref{sec:dataset} we perform extensive experiments using multiple images and filters to evaluate and compare the performance of each method under different conditions. 

\subsection{Self guided filter\label{sec:filter1}}

\begin{figure}[h!]
    \centering
    \includegraphics[trim=0 0 0 0, clip,width=\textwidth]{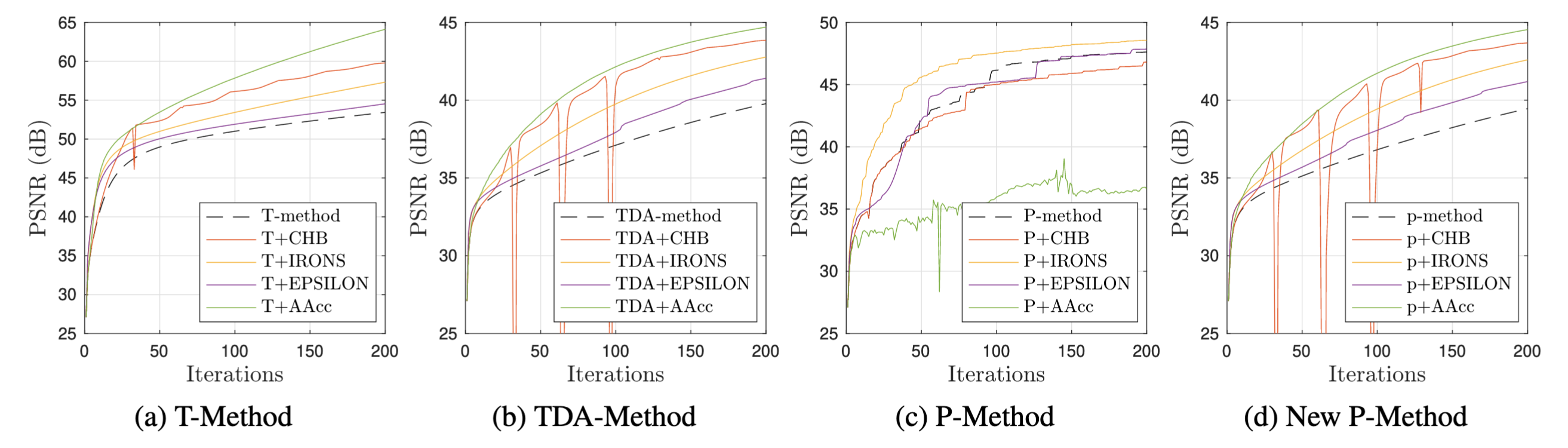}
    \caption{Reversing a Self Guided filter ($5\times5$ window size and $\epsilon=0.5$) with different reverse filtering methods and accelerated fixed point iteration.}
    \label{fig:AFP_PSNR}
\end{figure}

\begin{figure}[h!]
    \centering
\includegraphics[trim=0 0 0 0, clip,width=\textwidth]{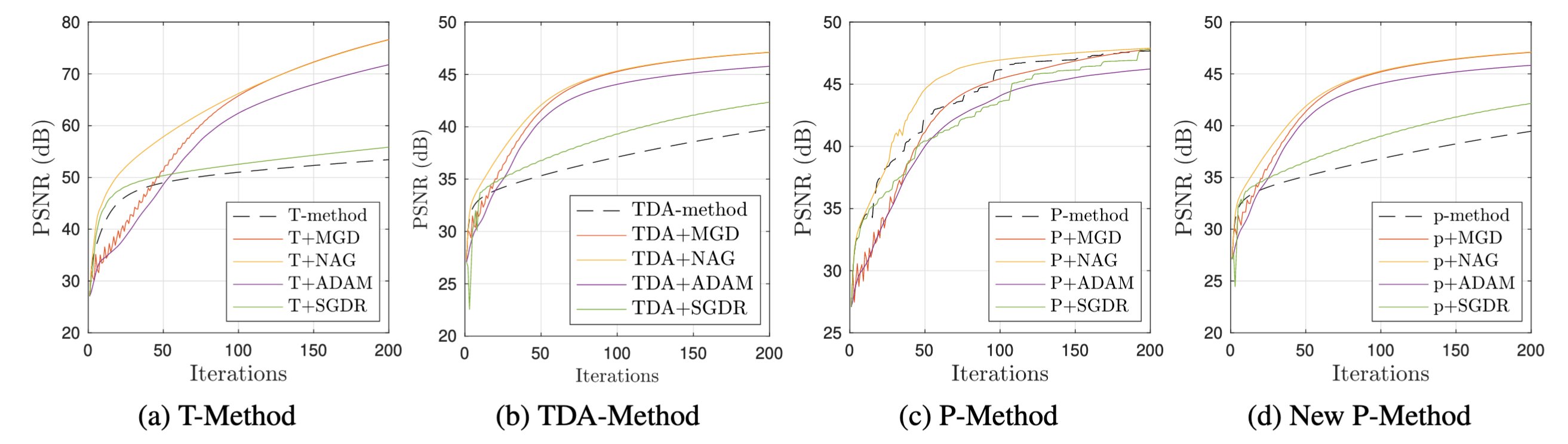}
    \caption{Reversing a Self Guided filter ($5\times5$ window size and $\epsilon=0.5$) with different reverse filtering methods and accelerated gradient descent.}
    \label{fig:AGD_PSNR}
\end{figure}

Here we use a Guided filter \cite{he2012guided} in a self guided configuration which is proven to be reversible by all the reverse image filtering methods under test here.  We slightly blur the famous \textit{cameraman.tif} image using a patch size of $5\times5$ and a $\epsilon = 0.05$. Then we recover the smoothed details by applying the T, TDA, P and p method with different acceleration schemes from FP and GD point of view. We recorded the PSNR values per iteration for each particular AFP and AGD method and present the result in Fig.\ref{fig:AFP_PSNR} and \ref{fig:AGD_PSNR} respectively. The curve for original methods (Without acceleration) are plotted with a dashed black line to simplify visualization.

Fig.\ref{fig:AFP_PSNR} shows that T, TDA and the p-method are improved with the acceleration from FP point o view. CHB, IRONS, EPSILON and AAcc produce a higher PSNR than the original methods. However, we can see that the P method is only improved by IRONS acceleration when reversing this particular filter. 

A similar observation can be made when applying gradient descent accelerations. Fig.\ref{fig:AGD_PSNR} shows that T, TDA and the p-method are improved with the acceleration. The increment on PSNR value achieved by AGD is higher than the increment produced by AFP. The P method shows a slight improvement in speed when using NAG but other methods are not successful.

AGD methods are better than AFP methods in this particular case. Also, AFP methods such as IRONS or EPSILON require more function calls than AGD methods, making AGD methods more appropriate to reverse this filter. 

\begin{figure}[h!]
    \centering
    \begin{subfigure}{0.24\linewidth}
        \includegraphics[trim=0 0 0 0, clip,width=\linewidth]{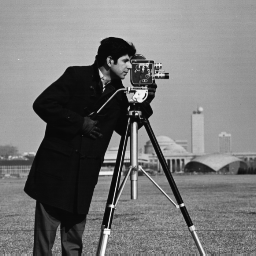}
        \caption{Ground Truth}
    \end{subfigure}
        \begin{subfigure}{0.24\linewidth}
        \includegraphics[trim=0 0 0 0, clip,width=\linewidth]{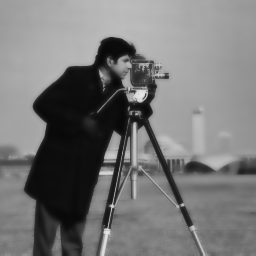}
        \caption{Input image}
    \end{subfigure}
        \begin{subfigure}{0.24\linewidth}
        \includegraphics[trim=0 0 0 0, clip,width=\linewidth]{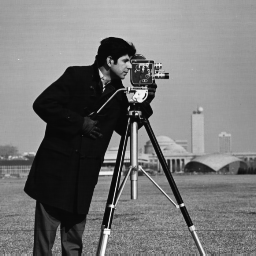}
        \caption{T-method}
    \end{subfigure}
        \begin{subfigure}{0.24\linewidth}
        \includegraphics[trim=0 0 0 0, clip,width=\linewidth]{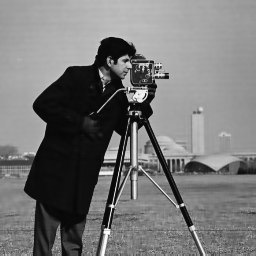}
        \caption{TDA-method}
    \end{subfigure}
        \begin{subfigure}{0.24\linewidth}
        \includegraphics[trim=0 0 0 0, clip,width=\linewidth]{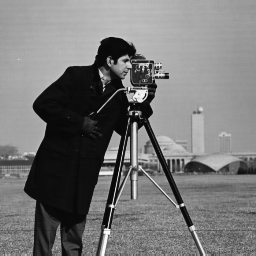}
        \caption{P-method}
    \end{subfigure}
            \begin{subfigure}{0.24\linewidth}
        \includegraphics[trim=0 0 0 0, clip,width=\linewidth]{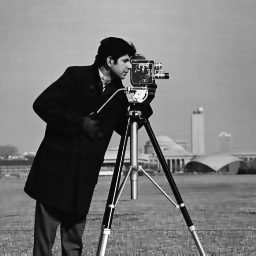}
        \caption{p-method}
    \end{subfigure}
   \begin{subfigure}{0.24\linewidth}
        \includegraphics[trim=0 0 0 0, clip,width=\linewidth]{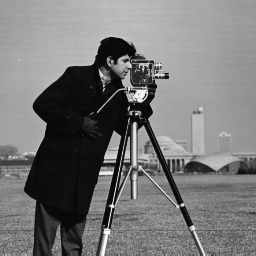}
        \caption{T+AFP}
    \end{subfigure}
     \begin{subfigure}{0.24\linewidth}
        \includegraphics[trim=0 0 0 0, clip,width=\linewidth]{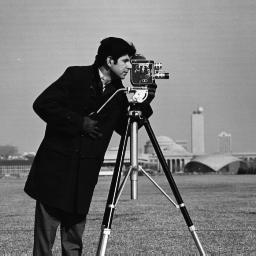}
        \caption{T+AGD}
    \end{subfigure}
    
    \caption{Reversing a Self Guided filter ($5\times5$ window size and $\epsilon=0.5$) with different reverse filtering methods. a) Ground truth image, b) Input blurred image, c) d) e) f) results using the classic T, TDA, P and p methods respectively without modification. g) Best result applying AFP which was achieved by the T+AA method, h) best result applying AGD achieved by T+NAG.}
    \label{fig:GF_visual}
\end{figure}

In Fig.\ref{fig:GF_visual} we can see the original image and its filtered input as well as the restored versions, we only display the result of the original methods and the best performer for each acceleration type.  We can see that the accelerated methods are able to recover the  details and edges smoothed by the self guided filter. 

\subsection{Motion Blur filter \label{sec:filter2}}

\begin{figure}[h!]
    \centering
\includegraphics[trim=0 0 0 0, clip,width=\textwidth]{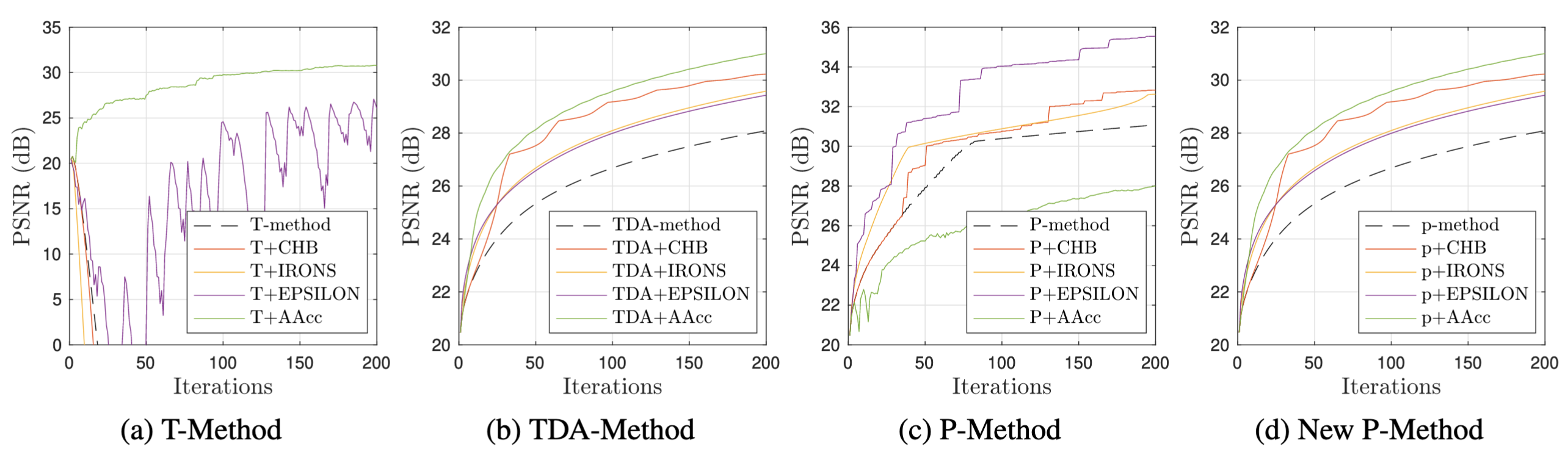}
    \caption{Reversing an MOTION filter with different reverse filtering methods and accelerated fixed point iteration.}
    \label{fig:AFP_PSNR2}
\end{figure}

\begin{figure}[h!]
    \centering
   \includegraphics[trim=0 0 0 0, clip,width=\textwidth]{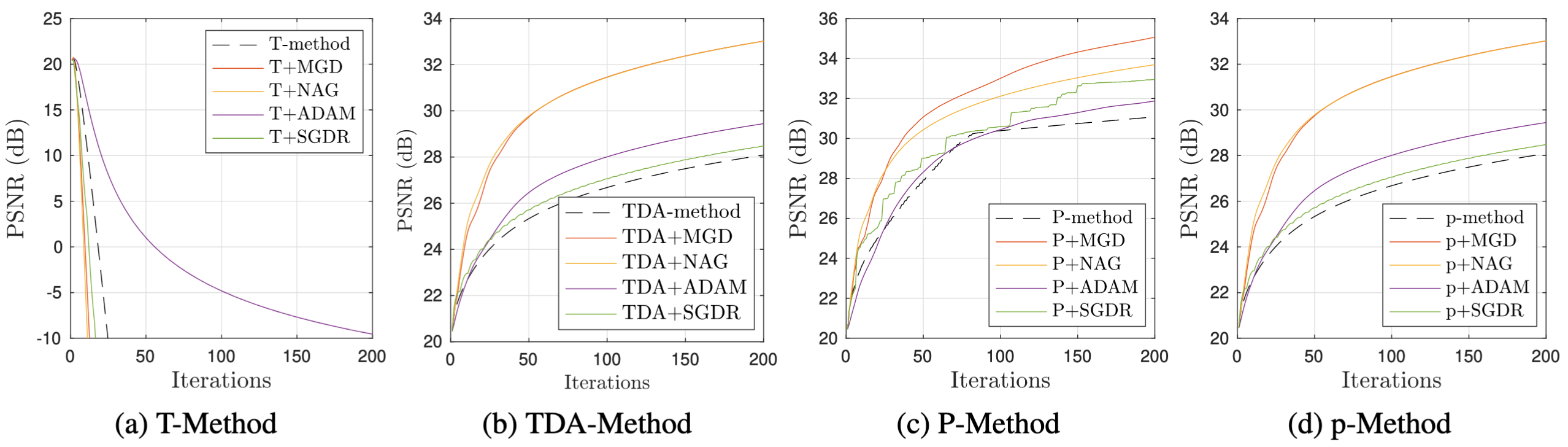}
    \caption{Reversing an MOTION with different reverse filtering methods and accelerated gradient descent.}
    \label{fig:AGD_PSNR2}
\end{figure}

In this section we use a Motion Blur Filter, which is a more challenging filter to reverse than the guided filter due to the higher degree of smoothing. The settings to generate the blurred image are $ l=20$, and $\theta = 45^{\circ}$ which produces a PSNR=20.45dB. Similar to previous section we split the results into FP and GD accelerations and display the results in Fig.\ref{fig:AFP_PSNR2} and \ref{fig:AGD_PSNR2}. 

The original T-method does not converge, meaning that is not able to reverse the effect of this filter since the PSNR decreases after each iteration. We can see that the only two acceleration methods able to help with this issue are the AAcc and EPSILON. AAcc shows an smooth increment of PSNR per iteration while the Epsilon acceleration show significant fluctuations. Fig.\ref{fig:AFP_PSNR2} and \ref{fig:AGD_PSNR2} also show that the remaining methods (TDA, P and p methods) are improved with FP and GD acceleration when reversing the Motion Blur Filter. Again we can see that AGD is more effective than AFP to improve these methods since the PSNR values achieved by GD methods are higher than the values achieved by FP.

\begin{figure}[h!]
    \centering
    \begin{subfigure}{0.24\linewidth}
        \includegraphics[trim=0 3cm 0 0, clip,width=\linewidth]{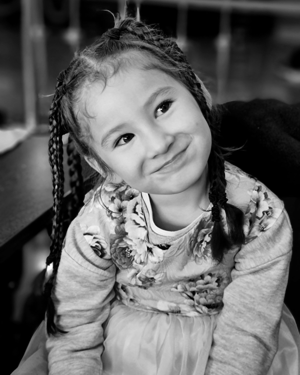}
        \caption{Ground Truth}
    \end{subfigure}
        \begin{subfigure}{0.24\linewidth}
        \includegraphics[trim=0 3cm 0 0, clip,width=\linewidth]{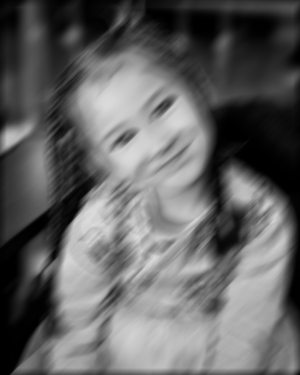}
        \caption{Input image}
    \end{subfigure}
        \begin{subfigure}{0.24\linewidth}
        \includegraphics[trim=0 3cm 0 0, clip,width=\linewidth]{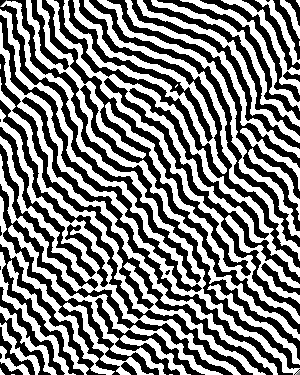}
        \caption{T-method}
    \end{subfigure}
        \begin{subfigure}{0.24\linewidth}
        \includegraphics[trim=0 3cm 0 0, clip,width=\linewidth]{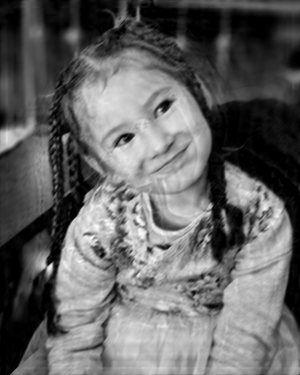}
        \caption{TDA-method}
    \end{subfigure}
        \begin{subfigure}{0.24\linewidth}
        \includegraphics[trim=0 3cm 0 0, clip,width=\linewidth]{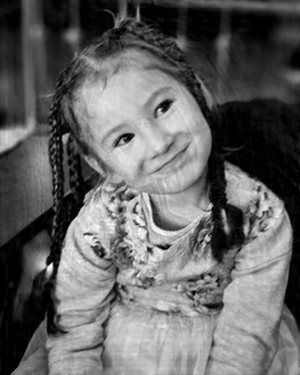}
        \caption{P-method}
    \end{subfigure}
            \begin{subfigure}{0.24\linewidth}
        \includegraphics[trim=0 3cm 0 0, clip,width=\linewidth]{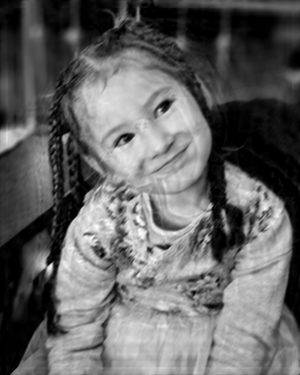}
        \caption{p-method}
    \end{subfigure}
   \begin{subfigure}{0.24\linewidth}
        \includegraphics[trim=0 3cm 0 0, clip,width=\linewidth]{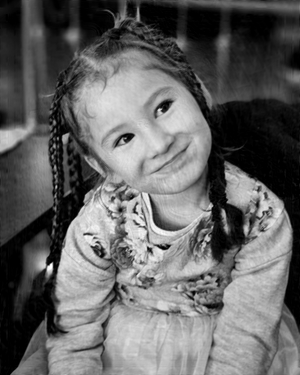}
        \caption{P+AFP}
    \end{subfigure}
     \begin{subfigure}{0.24\linewidth}
        \includegraphics[trim=0 3cm 0 0, clip,width=\linewidth]{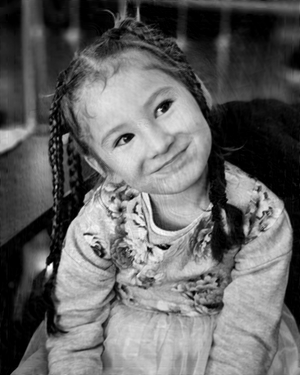}
        \caption{P+AGD}
    \end{subfigure}
    
    \caption{Reversing an MOTION filter with different reverse filtering methods. a) Ground truth image, b) Input blurred image, c) d) e) f) results using the classic T, TDA, P and p methods respectively without modification. g) Best result applying AFP which was achieved by the P+Epsilon method, h) best result applying AGD achieved by P+MGD.}
    \label{fig:MF_visual}
\end{figure}

A visual comparison is shown in Fig.\ref{fig:MF_visual}, we can see that the input image has heavily blurred details which can be challenging to recover. The T method does not converge so the result image is totally distorted after 200 iterations. TDA, P and P methods are able to recover most of the missing details, however some artifacts or distortion appeared on the girl's face.  Fig.\ref{fig:MF_visual}g and h present the best result for each acceleration type which are P+EPSILON and P+NAG. These results have significantly less distortion than the original methods which can be appreciated on the girl's face. We can see that applying acceleration significantly improve the result obtained with the same number of iterations. 

\begin{table}[h!]\label{tab:filters-to-be-reverse}
\centering
\caption{Parameter settings for the 14 filters}
\label{tab:filters_parameters}
\begin{tabular}{l l}
\toprule
\textbf{Filter} & \textbf{Parameters} \\
\midrule
Wiener Filter\footnote[2]{Matlab function \textit{wiener2.m} for 2-D adaptive noise-removal filtering} (WF)\cite{Gonzalez_DIP} &$w_{size} = 5 \times 5, \quad n=0.1 $ \\
Disk  Averaging Filter&$ r=3 $       \\
Motion Blur Filter &$ l=20,\quad  \theta = 45^{\circ}$   \\
Self Guided Gilter (GF)  \cite{he2012guided}     &$w_{size}= 5\times5, \quad  \epsilon=0.1$      \\
GF+Gauss.       &$w_{size}= 5\times5, \quad   \epsilon=0.1, \quad  \sigma = 5$     \\
Smoothing and Sharpening Filter (SSIF) \cite{deng2021guided} &$r=5, \quad \epsilon=0.1 ,\quad \kappa =0.1 ,\quad s = 1$ \\
WLS-based Decomposition for Multi-Scale Filtering  (WLS) \cite{farbman2008edge} &$\lambda=0.5, \alpha = 1.2$ \\
Structure Extraction from Texture via Relative Total Variation (RTV) \cite{xu2012structure}     &$\lambda=0.015, \quad \sigma=3, \quad \epsilon=0.05, \quad N_{iter}=1$     \\
Gaussian Filter (GS) \cite{Gonzalez_DIP}                                             & $ \sigma = 5$   \\
Adaptive Manifolds for Real-Time Filtering (AMF) \cite{gastal2012adaptive}  &$\sigma_s=20, \quad \sigma_r=0.4$  \\
Iterative Least Squares Filter (ILS) \cite{liu2020real} &$\lambda= 1,\quad p=0.8,\quad \epsilon=1\times10^{-4},\quad N_{iter}=4 $   \\
Image Smoothing via L0 minimization (L0) \cite{xu2011image} &$\lambda = 0.01,\quad \kappa = 2 $    \\
Bilateral Filter (BF) \cite{tomasi1998bilateral} &$ \epsilon=0.05,\quad \sigma_s=3 $  \\
Local Laplacian Filter (LLF) \cite{paris2011local} &$\sigma=0.2,\quad \alpha=10$      \\

\bottomrule
\end{tabular}

\end{table}

\begin{table}[h!]
\caption{Parameter settings for each reverse image filtering method and acceleration. Irons, Epsilon and Anderson acceleration are parameter free accelerations. }
\label{tab:settings}
\begin{tabular}{lllll}
\toprule
\textbf{Acceleration} & \multicolumn{1}{c}{\textbf{T}}   & \multicolumn{1}{c}{\textbf{TDA}} & \multicolumn{1}{c}{\textbf{P}}   & \multicolumn{1}{c}{\textbf{p}}   \\ 
\midrule
\textbf{None}         & $ \lambda = 1 $                  & $ \lambda = 1 $                  & $ \lambda = 1$                   & $ \lambda = 1 $                  \\
\textbf{NAG}          & $\alpha = 1, \beta = 0.9$        & $\alpha = 1, \beta = 0.9$        & $\alpha = 1, \beta = 0.9$        & $\alpha = 1, \beta = 0.9$        \\
\textbf{MGD}          & $\alpha = 1, \beta = 0.9$        & $\alpha = 1, \beta = 0.9$        & $\alpha = 1, \beta = 0.9$        & $\alpha = 1, \beta = 0.9$        \\
\textbf{ADAM}         & $ \beta_1 = 0.9 , \beta_2=0.999$ & $ \beta_1 = 0.9 , \beta_2=0.999$ & $ \beta_1 = 0.9 , \beta_2=0.999$ & $ \beta_1 = 0.9 , \beta_2=0.999$ \\
\textbf{SGDR} &
  \begin{tabular}[c]{@{}l@{}}$T = 5, \lambda_{min}=1, $\\ $\lambda_{max}=2$\end{tabular} &
  \begin{tabular}[c]{@{}l@{}}$T = 5, \lambda_{min}=0, $\\ $\lambda_{max}=3$\end{tabular} &
  \begin{tabular}[c]{@{}l@{}}$T = 5, \lambda_{min}=0,$\\ $\lambda_{max}=1$\end{tabular} &
  \begin{tabular}[c]{@{}l@{}}$T = 5, \lambda_{min}=0,$\\ $ \lambda_{max}=3$\end{tabular} \\
\textbf{CHB}          & $\alpha =3 , T=32$               & $\alpha = 3, T=32$               & $\alpha = 1, T=32$               & $\alpha = 3, T=32$               \\
\bottomrule
\end{tabular}
\end{table}

\subsection{Test on Dataset.\label{sec:dataset}}

The performance of a reverse image filtering method and its accelerations strongly depends on the filter being reversed and its settings, so to evaluate them rigorously we filtered 20 images from the BSD300 dataset \cite{martin2001database} and filter them with 14 commonly used image filters listed in table \ref{tab:filters_parameters}.The settings used for each reverse image filter method and acceleration are shown in table \ref{tab:settings}. We record the PSNR per iteration for each image and filter. The level of degradation depends on the filter used, so we re-scale the result to show the percentage of improvement using the following equation: 

\begin{equation}
\bar{p}_k^i= \frac{p_k^i - p_0^i}{p_0^i}\times 100
\end{equation}
where $p_k^i$ and $p_0^i$ represent respectively the PSNR value at the $k$th iteration and $0$th iteration when reversing the effect of a filter on an image $i$.

We then find the maximum percentage of increment achieved for each image and average the results as follows:

\begin{equation}
p_{max}=\frac{1}{20}\sum_{i=1}^{20}(\max(\bar{p}_k^i))
\end{equation}

We compute $p_{max}$ for each filter and reverse iamge filtering method and use it as a metric to evaluate their performance. Results are presented in Table \ref{tab:T_Dataset}-\ref{tab:P2_dataset}.

Main observations.
\begin{itemize}
\item All the methods are improved at some extent by at least one acceleration method. 
    \item AA is surprisingly performing well with the T-method when it is not convergent for certain filters such as WF, Disk and Motion.   
    \item The performance when reversing a few filters such as LO and GS is poor, since the average maximum improvement is around 10\%.
    \item The acceleration methods that best perform when using the P method are notably the IRONS and EPSILON algorithm. 
    \item Overall, the T method and its accelerated versions are the algorithms that can achieve higher levels of improvements compared with TDA, P and p based algorithms. 
    
\end{itemize}

\newpage


\begin{table}[h!]
\centering
\caption{T-method and its variants. }
\label{tab:T_Dataset}
\begin{tabular}{llllllllll}
\toprule
\textbf{Filter} & \textbf{T} & \textbf{NAG} & \textbf{MGD} & \textbf{ADAM} & \textbf{SGDR} & \textbf{IRONS} & \textbf{EPSILON} & \textbf{CHB} & \textbf{AA} \\
\midrule
\textbf{WF}     & 2.2           & 2.2            & 2.1           & 0.5   & 1.3           & 1.3           & 1.2           & 2.2   & \textbf{29.9}   \\
\textbf{Disk}   & 3.4           & 3.4            & 3.2           & 1.1   & 3.1           & 3.2           & 14.3          & 3.4   & \textbf{36.5}   \\
\textbf{Motion} & 2.2           & 2.2            & 2.2           & 0.7   & 1.3           & 1.3           & 12.6          & 2.2   & \textbf{36.3}   \\
\textbf{GF}     & 116.5         & \textbf{200.9} & 146.9         & 136.5 & 130.5         & 138.1         & 134.1         & 133.2 & 161.7           \\
\textbf{GF+GS}  & 28.5          & 49.1           & \textbf{49.2} & 33.8  & 32.3          & 34.2          & 31.6          & 33.0  & 40.5            \\
\textbf{SSIF}   & 1058.1        & 794.0          & 47.2          & 175.6 & 859.2         & 1057.4        & 1058.7        & 899.5 & \textbf{1059.4} \\
\textbf{WLS}    & 22.8          & 22.5           & 17.1          & 17.9  & \textbf{23.0} & 22.8          & 22.9          & 22.9  & 22.9            \\
\textbf{RTV}    & 35.8          & 29.8           & 21.0          & 27.4  & \textbf{35.9} & 35.8          & 35.8          & 35.2  & 35.9            \\
\textbf{GS}     & 8.4           & 8.5            & 7.1           & 4.1   & 8.4           & 8.4           & 8.3           & 8.4   & \textbf{10.2}   \\
\textbf{AMF}    & 98.4          & 91.2           & 74.7          & 87.7  & 99.1          & \textbf{99.7} & 99.0          & 95.7  & 99.3            \\
\textbf{ILS}    & 50.6          & \textbf{59.2}  & 56.7          & 49.5  & 52.7          & 53.7          & 52.8          & 52.9  & 58.7            \\
\textbf{L0}     & 8.7           & 6.0            & 5.4           & 6.0   & 6.4           & 8.7           & \textbf{11.2} & 7.8   & 10.8            \\
\textbf{BF}     & 97.5          & 88.3           & 63.8          & 67.7  & 88.3          & 100.2         & 99.0          & 93.6  & \textbf{105.4}  \\
\textbf{LLF}    & 38.0 & 32.5           & 19.0          & 25.0  & 33.1          & 38.0          & 38.0          & 38.0  & \textbf{38.1}        \\
\bottomrule
\end{tabular}
\end{table}

\begin{table}[h!]
\centering
\caption{TDA-method and its variants}
\label{tab:TDA_dataset}
\begin{tabular}{llllllllll}
\toprule
\textbf{Filter} & \textbf{TDA} & \textbf{NAG} & \textbf{MGD} & \textbf{ADAM} & \textbf{SGDR} & \textbf{IRONS} & \textbf{EPSILON} & \textbf{CHB} & \textbf{AA} \\
\midrule
\textbf{WF}     & 22.6  & \textbf{35.7} & 34.8          & 24.3          & 25.9  & 26.8          & 25.9          & 25.8  & 32.4           \\
\textbf{Disk}   & 19.7  & 38.6          & \textbf{38.7} & 27.4          & 24.6  & 25.4          & 25.3          & 24.3  & 31.5           \\
\textbf{Motion} & 24.3  & 43.6          & \textbf{43.7} & 29.6          & 29.0  & 29.8          & 30.6          & 28.6  & 35.8           \\
\textbf{GF}     & 51.4  & 78.5          & 77.9          & 73.7          & 54.8  & 59.8          & 60.7          & 58.1  & \textbf{78.9}  \\
\textbf{GF+GS}  & 9.6   & 16.7          & \textbf{16.7} & 13.4          & 11.2  & 11.5          & 10.7          & 11.1  & 13.9           \\
\textbf{SSIF}   & 412.7 & 266.2         & 33.0          & 162.7         & 373.7 & 771.7         & 752.9         & 470.7 & \textbf{936.5} \\
\textbf{WLS}    & 9.9   & 9.6           & 8.1           & \textbf{11.1} & 7.2   & 10.0          & 9.7           & 9.5   & 9.5            \\
\textbf{RTV}    & 14.9  & 13.5          & 7.8           & 11.7          & 14.3  & \textbf{16.5} & 15.3          & 12.7  & 15.8           \\
\textbf{GS}     & 6.4   & 8.9           & \textbf{8.9}  & 7.1           & 7.0   & 7.1           & 6.8           & 6.9   & 7.9            \\
\textbf{AMF}    & 34.6  & \textbf{57.7} & 52.3          & 46.2          & 39.6  & 41.4          & 37.5          & 35.7  & 39.3           \\
\textbf{ILS}    & 26.8  & \textbf{29.4} & 26.3          & 22.6          & 26.0  & 28.7          & 28.2          & 24.9  & 27.5           \\
\textbf{L0}     & 1.2   & 0.1           & 0.3           & 2.1           & 0.0   & 1.1           & \textbf{6.0}  & 0.9   & 5.2            \\
\textbf{BF}     & 33.5  & 43.6          & 52.8          & \textbf{58.1} & 28.0  & 41.1          & 33.2          & 31.0  & 36.6           \\
\textbf{LLF}    & 23.1  & 5.8           & 5.7           & 21.2          & 0.4   & 23.0          & \textbf{25.4} & 18.2  & 24.8             \\
\bottomrule
\end{tabular}
\end{table}


\begin{table}[h!]
\centering
\caption{P-method and its variants}
\label{tab:P_dataset}
\begin{tabular}{llllllllll}
\toprule
\textbf{Filter} & \textbf{P} & \textbf{NAG} & \textbf{MGD} & \textbf{ADAM} & \textbf{SGDR} & \textbf{IRONS} & \textbf{EPSILON} & \textbf{CHB} & \textbf{AA} \\
\midrule
\textbf{WF}     & 33.3 & 36.7          & 35.5 & 30.4          & 34.2 & \textbf{40.2} & 38.0          & 35.6 & 21.2 \\
\textbf{Disk}   & 30.3 & 41.4          & 44.9 & 37.1          & 36.1 & 38.4          & \textbf{46.6} & 39.2 & 18.5 \\
\textbf{Motion} & 32.8 & 46.5          & 50.1 & 40.8          & 43.8 & 40.3          & \textbf{51.7} & 43.7 & 23.0 \\
\textbf{GF}     & 81.1 & 82.4          & 78.9 & 77.3          & 75.0 & \textbf{95.7} & 77.2          & 76.2 & 45.0 \\
\textbf{GF+GS}  & 17.7 & 18.1          & 19.2 & 15.5          & 17.0 & \textbf{21.4} & 18.5          & 17.5 & 10.6 \\
\textbf{SSIF}    & 962.8      & 238.1        & 25.7         & 159.7         & 862.0         & \textbf{1053.4} & 1047.1           & 909.4        & 561.0       \\
\textbf{WLS}    & 10.9 & 11.0          & 7.9  & \textbf{12.0} & 11.2 & 11.1          & 11.1          & 10.9 & 10.7 \\
\textbf{RTV}    & 18.3 & 14.5          & 6.8  & 13.6          & 18.7 & \textbf{20.6} & 20.5          & 19.3 & 13.2 \\
\textbf{GS}     & 9.4  & 9.5           & 10.3 & 8.1           & 8.8  & \textbf{11.5} & 10.8          & 9.6  & 6.6  \\
\textbf{AMF}    & 51.0 & \textbf{60.0} & 42.6 & 50.9          & 49.6 & 59.1          & 51.5          & 51.4 & 33.0 \\
\textbf{ILS}    & 29.1 & 28.9          & 25.6 & 25.2          & 29.1 & \textbf{31.6} & 30.3          & 29.1 & 27.0 \\
\textbf{L0}     & 5.6  & 2.0           & 1.3  & 2.6           & 5.5  & 5.8           & \textbf{6.1}  & 5.7  & 5.7  \\
\textbf{BF}     & 38.5 & 57.6          & 55.1 & \textbf{60.8} & 42.1 & 56.1          & 50.6          & 42.3 & 24.9 \\
\textbf{LLF}    & 23.3 & 6.8           & 4.9  & 23.7          & 22.7 & 23.2          & \textbf{24.6} & 23.6 & 24.5 \\
\bottomrule
\end{tabular}
\end{table}

\begin{table}[h!]
\centering
\caption{p-method and its variants. }

\label{tab:P2_dataset}
\begin{tabular}{llllllllll}
\toprule
\textbf{Filter} & \textbf{p} & \textbf{NAG} & \textbf{MGD} & \textbf{ADAM} & \textbf{SGDR} & \textbf{IRONS} & \textbf{EPSILON} & \textbf{CHB} & \textbf{AA} \\
\midrule
\textbf{WF}     & 22.6  & \textbf{35.6} & 34.4          & 24.2          & 25.9  & 26.8          & 25.9         & 25.9  & 32.4           \\
\textbf{Disk}   & 19.7  & 38.6          & \textbf{38.7} & 27.4          & 24.6  & 25.4          & 25.3         & 24.3  & 31.5           \\
\textbf{Motion} & 24.3  & 43.6          & \textbf{43.7} & 29.6          & 29.0  & 29.8          & 30.6         & 28.6  & 35.8           \\
\textbf{GF}     & 50.6  & 78.3          & 77.7          & 74.1          & 55.9  & 59.3          & 59.1         & 57.5  & \textbf{78.4}  \\
\textbf{GF+GS}  & 9.6   & 16.7          & \textbf{16.7} & 13.4          & 11.2  & 11.5          & 10.7         & 11.1  & 13.9           \\
\textbf{SSIF}   & 412.5 & 271.5         & 36.0          & 163.1         & 383.9 & 771.7         & 755.3        & 476.4 & \textbf{948.0} \\
\textbf{WLS}    & 10.2  & \textbf{11.6} & 9.8           & 11.5          & 10.6  & 10.5          & 10.0         & 9.8   & 9.9            \\
\textbf{RTV}    & 14.3  & 15.1          & 11.2          & 13.2          & 15.5  & \textbf{16.0} & 14.4         & 14.0  & 15.9           \\
\textbf{GS}     & 6.4   & 8.9           & \textbf{8.9}  & 7.1           & 7.0   & 7.1           & 6.8          & 6.9   & 7.9            \\
\textbf{AMF}    & 34.7  & \textbf{58.7} & 53.2          & 46.9          & 40.3  & 42.0          & 37.6         & 35.9  & 40.8           \\
\textbf{ILS}    & 26.1  & \textbf{29.0} & 26.1          & 22.6          & 26.9  & 28.2          & 27.5         & 25.9  & 27.2           \\
\textbf{L0}     & 3.4   & 1.3           & 1.6           & 2.8           & 0.5   & 3.3           & \textbf{6.1} & 2.7   & 5.4            \\
\textbf{BF}     & 33.1  & 51.7          & 54.9          & \textbf{58.2} & 32.3  & 41.2          & 32.1         & 32.0  & 35.9           \\
\textbf{LLF}    & 25.9  & 10.5          & 7.2           & 24.1          & 6.8   & 25.9          & 26.2         & 19.6  & \textbf{26.9}         \\
\bottomrule

\end{tabular}
\end{table}

To complete this section, we present a summary in Table \ref{tab:summary_dataset} that clearly shows which reverse filtering and acceleration method work better for each filter. We can see that when the T method converges, its accelerated versions achieve higher PSNR improvement than TDA, P and p based methods. For those filters, when the T method does not converge, AA is the most appropriate acceleration to use since it makes the T method converge, however the PSNR improvement values achieved by P method based methods are higher. 

\begin{table}[h!]
\centering
\caption{Result summary. Numbers presented in red italic font represents a non-convergent result. Numbers presented in black bold font represent the best result for the particular filter. }
\label{tab:summary_dataset}
\begin{tabular}{|l|ll|ll|ll|ll|}
\toprule
\textbf{Filter} & \multicolumn{2}{c|}{\textbf{T}} & \multicolumn{2}{c|}{\textbf{TDA}} & \multicolumn{2}{c|}{\textbf{P}} & \multicolumn{2}{c|}{\textbf{p}} \\
\midrule
\textbf{WF}     & 29.9            & AA      & 35.7  & NAG     & \textbf{40.2} & IRONS   & 35.6  & NAG     \\
\textbf{Disk}   & 36.5            & AA      & 38.7  & MGD     & \textbf{46.6} & EPSILON & 38.7  & MGD     \\
\textbf{Motion} & 36.3            & AA      & 43.7  & MGD     & \textbf{51.7} & EPSILON & 43.7  & MGD     \\
\textbf{GF}     & \textbf{200.9}  & NAG     & 78.9  & AA      & 95.7          & IRONS   & 78.4  & AA      \\
\textbf{GF+GS}  & \textbf{49.2}   & MGD     & 16.7  & MGD     & 21.4          & IRONS   & 16.7  & MGD     \\
\textbf{SSIF}   & \textbf{1059.4} & AA      & 936.5 & AA      & 1053.4        & IRONS   & 948.0 & AA      \\
\textbf{WLS}    & \textbf{23.0}   & SGDR    & 11.1  & ADAM    & 12.0          & ADAM    & 11.6  & NAG     \\
\textbf{RTV}    & \textbf{35.9}   & SGDR    & 16.5  & IRONS   & 20.6          & IRONS   & 16.0  & IRONS   \\
\textbf{GS}     & 10.2            & AA      & 8.9   & MGD     & \textbf{11.5} & IRONS   & 8.9   & MGD     \\
\textbf{AMF}    & \textbf{99.7}   & IRONS   & 57.7  & NAG     & 60.0          & NAG     & 58.7  & NAG     \\
\textbf{ILS}    & \textbf{59.2}   & NAG     & 29.4  & NAG     & 31.6          & IRONS   & 29.0  & NAG     \\
\textbf{L0}     & \textbf{11.2}   & EPSILON & 6.0   & EPSILON & 6.1          & EPSILON & 6.1   & EPSILON \\
\textbf{BF}     & \textbf{105.4}  & AA      & 58.1  & ADAM    & 60.8          & ADAM    & 58.2  & ADAM    \\
\textbf{LLF}    & \textbf{38.1}   & AA      & 25.4  & EPSILON & 24.6          & EPSILON & 26.9  & AA \\
\bottomrule
\end{tabular}
\end{table}

\section{Conclusion}

In this paper, we first demonstrated successfully that the T, P and TDA method can be formulated as both fixed point iteration and gradient descent optimization, which leads to new algorithms and insights. Also, after compiling and modifying a set of existing methods to accelerate fixed-point iterations and gradient descent, we showed their applicability to reverse image filtering. As an example, we reversed the effect of a self guided filter and a Motion blur filter. In both cases, most of the acceleration methods outperformed the original method.

 An extensive experiment using a filter bank and an image dataset showed that for every reverse image filter method there is at least one acceleration method that improve its convergence speed and in some cases such as Anderson acceleration applied to the T-method the performance improves significantly, the accelerated method can converge even when reversing filters where the original method failed.

\appendix

\section*{Appendix A}
The T-method, which is originally called zero-order reverse filter \cite{Tmethod}, can be formulated as a fixed point iteration problem. This is shown in section \ref{sec:fp-T-method}. It is computationally most efficient among all reverse filtering methods considered in this paper. It produces good results when the filter to be reversed satisfies contraction mapping condition is satisfied. Otherwise the T-method iteration is unbounded. The stability of the T-method in reversing a linear low pass filter is analysed in our previous work \cite{TDAmethod}. The T-method can be also be motivated from Bregman's iteration point view \cite{Deng_EL2019} which proves that when a blurring filter is the result of solving a variational problem, then the T-method can completely recover the original image.

The R-method, which is originally called the rendition algorithm \cite{Rmethod}, is developed by using gradient descent to minimize a cost function. Since the filter is unknown, approximations/assumptions about the gradient were introduced. The convergence condition for the R-method is that the filter function $g(.)$ must be Lipschitz continuous \cite{biLipschitz}. The R-method can reverse the effect of a wider range of operators than the T-method. Due to the assumptions and approximations, the R-method can only reverse the effects of filters which mildly alter the original image. In a recent paper \cite{polyBlur}, similar ideas as that of the R- and T-method are used to develop an algorithm to remove mild defocus and motion blur from natural images. 

The F-method, which is formulated in the frequency domain \cite{Fmethod}, uses the Newton-Raphson technique to invert unknown linear filters. Approximation of the gradient is also used. Although the F-method produces good results for some filters and is reported to have better performance than the T-method and the R-method, it only deals with LSI filters. If the frequency response of the LSI filter has zeros in the frequency range of $[0,\pi]$ then the iteration is unstable. This is a classical problem in inverse filtering, which can be dealt with by using a Wiener filter in the Fourier domain.

The P-method and the S-method \cite{Pmethod} are inspired by gradient descent methods studied by Polak \cite{polyak1969minimization} and Steffensen \cite{steffensen1933remarks}. Both methods have been demonstrated to produce good results in reversing the effects of a larger number of linear and non-linear filters than the T-method and the F-method. A limitation is their inherent high computational costs of $O(n^2)$ due to the calculation of the 2-norm of a matrix in each iteration. Because the P-method is more stable than the S-method \cite{Pmethod}, we only consider the P-method in this work.

The TDA-method \cite{TDAmethod} is based on a patched-based approximation of local gradient using the total derivative. It is as computationally efficient as the T-method in that they are at the same level of complexity. It is also as effective as the P-method in that they can reverse almost the same set of filters. Thus, the TDA-method achieves a good trade-off between efficiency and effectiveness.

\end{document}